\documentclass[submission,copyright,creativecommons]{eptcs}
\providecommand{\event}{PLACES 2014} 

\usepackage{amssymb}
\usepackage{amsmath}
\usepackage{proof}
\usepackage{arydshln}
\usepackage{color}

\newtheorem{definition}{Def.}
\newtheorem{lemma}{Lemma}

\newtheorem{theorem}{Theorem}
\newcommand{\pic}{$\pi$-calculus}
\newcommand{\respi}{$\mathit{ReS}\pi$}
\newcommand{\respiBold}{{\bf\textit{ReS}$\pi$}}

\newcommand{\ifPi}{\texttt{if}}
\newcommand{\thenPi}{\texttt{then}}
\newcommand{\elsePi}{\texttt{else}}

\newcommand{\ce}[1]{\bar{#1}}
\newcommand{\requestAct}[3]{\ce{#1}(#2) . #3}
\newcommand{\acceptAct}[3]{#1(#2) . #3}
\newcommand{\requestPrefix}[2]{\overline{#1}(#2)}
\newcommand{\acceptPrefix}[2]{#1(#2)}
\newcommand{\send}[2]{#1!\langle#2\rangle}
\newcommand{\sendAct}[3]{\send{#1}{#2} .  #3}
\newcommand{\receive}[2]{#1?(#2)}
\newcommand{\receiveAct}[3]{\receive{#1}{#2} . #3}
\newcommand{\select}[2]{#1 \triangleleft #2}
\newcommand{\selectAct}[3]{\select{#1}{#2} . #3}
\newcommand{\branching}[1]{#1 \triangleright}
\newcommand{\branch}[2]{#1\, :\, #2}
\newcommand{\branchSep}{,}
\newcommand{\branchAct}[2]{\branching{#1} \{#2\}}

\newcommand{\ifthenelseAct}[3]{\ifPi\ #1\ \thenPi\ #2\ \elsePi\ #3}
\newcommand{\inact}{\mathbf{0}}
\newcommand{\res}[1]{(\nu #1)}

\newcommand{\recAct}[2]{\mu #1.#2}

\newcommand{\ctrue}{\texttt{true}}
\newcommand{\cfalse}{\texttt{false}}
\newcommand{\commit}[1]{\texttt{commit}(#1)}

\newcommand{\subst}[2]{[ #1 / #2 ]}
\newcommand{\msubst}[1]{[ #1 ]}
\newcommand{\ssubst}[2]{#1 / #2}

\newcommand{\red}{\rightarrow}

\newcommand{\ptag}[1]{#1 :}
\newcommand{\memBegin}{\langle}
\newcommand{\memEnd}{\rangle}
\newcommand{\amem}[3]{\memBegin #1,#2,#3\memEnd}
\newcommand{\cmem}[3]{\memBegin #1,#2,#3\memEnd}
\newcommand{\fmem}[1]{\memBegin #1\memEnd}

\newcommand{\forkev}{\rightrightarrows}
\newcommand{\fev}[3]{#1\forkev (#2,#3)}
\newcommand{\actEv}[3]{#1-\!\!#2\!\!\rightarrow#3}
\newcommand{\storedAct}{\mathrm{A}}
\newcommand{\initAct}[6]{#1(#2)(#3)\res{#6}#4#5}
\newcommand{\comAct}[5]{#1\langle#2\rangle(#3)#4#5}

\newcommand{\selAct}[4]{\select{#1}{#2}\,#3 \{#4\}}

\newcommand{\ifEv}[3]{#1 ?\, #2 \!:\! #3}
\newcommand{\nil}{\mathrm{nil}}

\newcommand{\fwred}{\twoheadrightarrow}
\newcommand{\bwred}{\rightsquigarrow}
\newcommand{\fwbwred}{\rightarrowtail}
\newcommand{\forgetMap}{\phi}
\newcommand{\forget}[1]{\phi(#1)}

\newcommand{\trace}{\sigma}

\newcommand{\causallyEq}{\asymp}

\newcommand{\intType}{\mathsf{int}}
\newcommand{\sharedChanType}[1]{\langle#1\rangle}
\newcommand{\outType}[1]{![#1]}
\newcommand{\inpType}[1]{?[#1]}

\newcommand{\branchType}[1]{\&[#1]}
\newcommand{\inactType}{\mathsf{end}}

\newcommand{\typing}{\Delta}

\newcommand{\comp}{\cdot}

\title{Towards Reversible Sessions\thanks{This work has been partially supported by 
the COST Action BETTY (IC1201), by the EU project ASCENS (257414), 
and by the Italian MIUR PRIN project CINA (2010LHT4KM).}}

\author{
Francesco Tiezzi
\institute{IMT Institute for Advanced Studies, Lucca, Italy}
\email{francesco.tiezzi@imtlucca.it}
\and
Nobuko Yoshida 
\institute{Imperial College, London, U.K.}
\email{n.yoshida@imperial.ac.uk}
}

\def\titlerunning{Towards Reversible Sessions}
\def\authorrunning{Tiezzi F. and Yoshida N.}
\begin{document}

\maketitle

\begin{abstract}
In this work, we incorporate reversibility into structured communication-based programming, to allow parties of a session to automatically undo, in a rollback fashion, the effect of previously executed interactions. This permits taking different computation paths along the same session, as well as reverting the whole session and starting a new one. 
Our aim is to define a theoretical basis for examining the interplay in concurrent systems between reversible computation and session-based interaction. We thus enrich a session-based variant of \pic\ with memory devices, dedicated to keep track of the computation history of sessions in order to reverse it.
We discuss our initial investigation concerning the definition of a session type discipline for the proposed reversible calculus, and its practical advantages for static verification of safe composition in communication-centric distributed software performing reversible computations.
\end{abstract}

\section{Introduction}
\label{sec:intro}

\emph{Reversible computing} aims at providing a computational model that, besides 
the standard (forward) executions,  also permits backward execution steps in order to 
undo the effect of previously performed forward computations.
Reversibility is a key ingredient in different application domains since many years 
and, recently, also in the design of reliable concurrent systems,
as it permits understanding existing patterns for programming reliable systems
(e.g., compensations, checkpointing, transactions) and, possibly, improving them or 
developing new ones.

A promising line of research on this topic advocates reversible variants of 
well-established process calculi, such as CCS and \pic, as formalisms for 
studying reversibility mechanisms in concurrent systems. 
Our work incorporates reversibility into a variant of \pic\ equipped 
with \emph{session} primitives supporting structured communication-based 
programming. 
A (binary) session consists in a series of reciprocal interactions 
between two parties, possibly with branching and recursion. Interactions on 
a session are performed via a dedicated private channel, which is generated 
when initiating the session.
Session primitives come together with a session type discipline offering a simple 
static checking framework to guarantee the correctness of communication patterns.

Practically, combining reversibility and sessions paves the way for the development of 
session-based communication-centric distributed software intrinsically capable of performing 
reversible computations. In this way, without further coding effort by the 
application programmer, the interaction among session parties is relaxed so that, e.g.,  
the computation can automatically go back, thus allowing to take different paths
when the current one is not satisfactory.  
As an application example, used in this paper for illustrating our approach, 
we consider a simple scenario involving 
a client and multiple providers offering 
the same service (e.g., on-demand video streaming). 
The client connects to a provider to request a given service  
(specifying, e.g., title of a movie, video quality, etc.). 
The provider replies with a quote determined according to the requested 
quality of service and to the servers status (current load, available bandwidth, etc.). 
Then, the client can either accept, negotiate or reject the quote.
If a problem occurs during the interaction between the client and the provider,
the computation can be reverted, in order to allow 
the client to automatically start a new session with (possibly) another provider.

The proposed reversible session-based calculus relies on memories to store information about 
interactions and their effects on the system, which otherwise would be lost during forward 
computations. This data is used to enable backward computations that revert the effects of 
the corresponding forward ones. Each memory is devoted to record data concerning a single 
event, which can correspond to the taking place of a communication action, a choice or a thread 
forking. Memories are connected each other, in order to keep track of the computation history, 
by using unique thread identifiers as links.
Like all other formalisms for reversible computing in concurrent settings, 
forward computations are undone in a \emph{causal-consistent} fashion, 
i.e.~backtracking does not have to necessarily follow the exact order of forward 
computations in reverse, because independent actions can be undone in a different order. 

The resulting formalism offers a theoretical basis for examining the interplay between reversible 
computations and session-based structured interactions.  
We notice that reversibility enables session parties not only 
to partially undo the interactions performed along the current session, but also to automatically undo 
the whole session and restart it, possibly involving different parties. The advantage of the reversible 
approach is that this behaviour is realised without explicitly implementing loops. 
On the other hand, the session type discipline affects reversibility as it forces concurrent interactions 
to follow structured communication patterns. 
In fact, linearizing behaviours on sessions reduces the effect of causal consistency, because 
concurrent interactions along the same session are forbidden and, hence, the rollback along a session 
follows a single path. However, interactions along different sessions are still concurrent and, therefore, they
can be reverted as usual in a causal-consistent fashion. 
Notably, interesting issues concerning reversibility and session types are still open questions, 
especially for what concerns the validity in the reversible setting of standard properties 
(e.g., progress enforcement) and possibly new properties (e.g., reversibility of ongoing session history, 
irreversible closure of sessions). 
\section{Related work}
\label{sect:relatedWork}
We review here the most closely related works, which concern the definition of 
reversible process calculi; we refer to \cite{TR} for a more detailed review of 
reversible calculi.

Reversible CCS (RCCS)~\cite{DanosK04} is the first proposal of reversible calculus, from 
which all subsequent works drew inspiration. 
To each currently running thread is associated an individual memory stack keeping track 
of past actions, as well as forks and synchronisations. Information pushed on the memory stacks,
upon doing a forward transition, can be then used for a roll-back.
The memories also serve as a naming scheme and yield unique identifiers for threads.
When a process divides in two sub-threads, each sub-thread inherits the father memory together 
with a fork number indicating which of the two sons the thread is. A drawback of this approach 
is that the parallel operator does not satisfy usual structural congruence rules as commutativity, 
associativity and nil process as neutral element. 

CCS-R~\cite{DanosK07a} is another reversible variant of CCS,
which mainly aims at formalising biological systems.  
Like RCCS, it relies on memory stacks for storing data needed for 
backtracking, which now also includes events corresponding to unfolding of process 
definitions. Differently from RCCS, specific identifiers are used to label threads, and 
a different approach is used for dealing with forking.

CCS with communication Keys (CCSK)~\cite{PhillipsU07} is a reversible process calculus
obtained by applying a general procedure to produce reversible calculi.
A relevant aspect of this approach is that it does not rely on memories for supporting backtracking. 
The idea is to maintain the structure of processes fixed throughout computations, thus
avoiding to consume guards and alternative choices. To properly revert synchronisations, 
the two threads have to agree on a key, uniquely identifying that communication. 

$\rho\pi$~\cite{LaneseMS10} is a reversible variant of the higher-order \pic.
It borrows from RCCS the use of memories for keeping track of past actions, 
although in $\rho\pi$ they are not stacks syntactically associated to threads,
but parallel terms each one dedicated to a single communication. 
The connection between memories and threads is kept by resorting to 
identifiers, which resemble CCSK keys. Fork handling is based on structured 
tags, used to connect the identifier of a thread with the identifiers of its sub-threads. 
This approach to reversibility has been applied in \cite{GLMT} to a
distributed tuple-based language.

Another reversible variant of \pic\ is R$\pi$~\cite{CristescuKV13}. 
Similarly to RCCS,  this calculus relies on memory stacks, now recording 
communication events and forking.
Differently from $\rho\pi$, it considers standard \pic\ (without choice and replication) 
as a host calculus and its semantics is defined in terms of a labelled transition relation. 

Finally, reversible structures~\cite{CardelliL11} is a simple computational calculus
for modelling chemical systems. Reversible structures does not exploit memories, but 
maintains the structure of terms and uses a special symbol to indicate the next forward 
and backward operations that a term can perform. 

In our work, we mainly take inspiration from the $\rho\pi$ approach.
In fact, all other approaches are based on CCS and  cannot be 
directly applied to a calculus with name-passing. Moreover, 
the $\rho\pi$ approach is preferable to the R$\pi$ one because the former proposes a 
reduction semantics, which we are interested in, while 
the latter proposes a labelled semantics, which would complicate our theoretical
framework in order to properly deal with scope extrusion. 
\section{Reversible Session-based \pic}
\label{sec:respi}

In this section, we introduce a reversible extension of a 
\pic\ enriched with primitives for managing \emph{binary} sessions,
i.e.~structured interactions between two parties.
We call \respi\ (\emph{Reversible Session-based} \pic) this formalism.
Due to lack of space, some technical details 
about semantics and results have been omitted; 
we refer the interested reader to \cite{TR} for a 
more complete account.

\paragraph{From \pic\ to \respi.}
Our approach to keep track of computation history in \respi\ is as follows: 
we tag processes with unique identifiers (tagged processes are called \emph{threads}) 
and use memories to store the information needed to reverse each single forward reduction. Thus, 
the history of a reduction sequence is stored in a number of small memories 
connected each other by using tags as links.  
In this way, \respi\ terms can perform, besides \emph{forward computations}
(denoted by $\fwred$), also  \emph{backward computations} (denoted by $\bwred$)
that undo the effect of the former ones in a causal-consistent fashion.

\pic\ \emph{processes} and \emph{expressions} are given by the grammars in Figure~\ref{fig:syntax_pi}.
\begin{figure}[t]
	\begin{centering}
	\begin{tabular}{@{\hspace*{-.1cm}}c@{\hspace*{-.1cm}}}
	{\small
	\begin{tabular}{@{}l@{\hspace*{.3cm}}l@{\hspace*{.3cm}}l@{\hspace*{.3cm}}l@{}}
	Shared channels $a$, $b$, \ldots
	&
	Session channels $s$, $s'$, \ldots
	&
	Session endpoints $s$, $\ce{s}$, $s'$, $\ce{s}'$, \ldots
	&
	Variables $x$, $y$, \ldots
	\\[.1cm]
	Labels $l$, $l'$, \ldots
	&
	Process variables $X$, $Y$, \ldots	
	&
	Tags $t$, $t'$, \ldots
	&
	Shared ids $u$ \ ::= \ $a \ \ \mid\ \ x$
	\\[.1cm]
	Channels $c$ \ ::= \ $a \ \ \mid\ \ s$
	&
	Names $h$  \ ::= \ $c \ \ \mid\ \ t$
	&
	\multicolumn{2}{@{}l@{}}{Session ids $k$ \ ::= \ $s \ \ \mid\ \ \ce{s} \ \ \mid\ \ x$}
	\\[.2cm]
	\end{tabular}}
	\\
	\hdashline
	\\[-.2cm]
	\begin{tabular}{@{}r@{\ \ }c@{\ \ }l@{}}
	\textbf{Processes}\ \
	$P$ & ::= & $\requestAct{u}{x}{P}
	\ \ \mid\ \
	\acceptAct{u}{x}{P}
	\ \ \mid\ \
	\sendAct{k}{e}{P}
	\ \ \mid\ \
	\receiveAct{k}{x}{P}
	\ \ \mid\ \
	\selectAct{k}{l}{P}
	\ \ \mid\ \
	\branchAct{k}{\branch{l_1}{P_1} \branchSep \ldots \branchSep \branch{l_n}{P_n}}$ \\
	& $\mid$ & $\ifthenelseAct{e}{P}{Q}
	\ \ \mid\ \
	P \mid Q
	\ \ \mid\ \
	\res{c}\, P
	\ \ \mid\ \
	X
	\ \ \mid\ \
	\recAct{X}{P}
	\ \ \mid\ \
	\inact$ 
	\\[.2cm]
	\textbf{Expressions}\ \
	$e$ & ::= & $v \ \ \mid\ \ x \ \ \mid\ \ \text{op}(e_1,\ldots,e_n)$  
	\\[.2cm]
	\textbf{Values}\ \ 
	$v$ & ::= & $\ctrue  \ \ \mid\ \ \cfalse  \ \ \mid\ \ 0,1,\ldots \ \mid \ \ a\ \mid\ \ s \ \ \mid\ \ \ce{s}$
	\\[.2cm]
	\textbf{\respiBold\ processes}\ \
	$M$ & ::= & $\ptag{t}P
	\ \ \mid\ \
	\res{h}\, M
	\ \ \mid\ \
	M \mid N
	\ \ \mid\ \
	m
	\ \ \mid\ \
	\nil$
	\\[.2cm]	
	\textbf{Memories}\ \
	$m$ & ::= & $\amem{\actEv{t_1}{\storedAct}{t_2}}{t_1'}{t_2'}
	\ \ \mid\ \
	\cmem{t}{\ifEv{e}{P}{Q}}{t'}
	\ \ \mid\ \
	\fmem{\fev{t}{t_1}{t_2}}$ 
	\\[.2cm]	
	$\storedAct$ & ::= & 	$\initAct{a}{x}{y}{P}{Q}{s}$ 
	\ \ $\mid$ \ \ $\comAct{k}{e}{x}{P}{Q}$ 
	\ \ $\mid$ \ \ $\selAct{k}{l_i}{P}{\branch{l_1}{P_1} \branchSep \ldots \branchSep \branch{l_n}{P_n}}$ 
	\\[.2cm]
	\end{tabular}	
	\\
	\hline
	\end{tabular}		
	\vspace*{-.2cm}
	\caption{\respi\ syntax}
	\label{fig:syntax_pi}
	\label{fig:syntax_respi}	
	\vspace*{-.2cm}
	\end{centering}
\end{figure}
The synchronisation on a shared channel $a$ 
of processes $\requestAct{a}{x}{P}$ and $\acceptAct{a}{y}{Q}$
initiates a session along a fresh session channel~$s$. 
This channel consists in a pair of (dual) endpoints, denoted by $s$ and $\ce{s}$ (such that
$\ce{\ce{s}}=s$), each one dedicated to one party to exchange values with the other.
These endpoints replace variables $x$ and $y$, by means of a substitution
application,  in order to be used by $P$ and $Q$, respectively, for later communications. 
Primitives $\sendAct{k}{e}{P}$ and $\receiveAct{k'}{x}{Q}$ 
denote output and input via session endpoints identified by $k$ and $k'$, respectively. 
These communication primitives realise the standard synchronous message passing, 
where messages result from expressions evaluation and may contain endpoints 
(\emph{delegation}). 
Constructs $\selectAct{k}{l}{P}$ and 
$\branchAct{k'}{\branch{l_1}{P_1}\branchSep\ldots\branchSep \branch{l_n}{P_n}}$
denote label selection and branching (with $l_1$, \ldots, $l_n$ 
pairwise distinct) via $k$ and $k'$, respectively. 
The above interaction primitives are combined by 
conditional choice, parallel composition, restriction, 
recursion and inaction.

\respi\ processes are built upon \pic\ processes by labelling them 
with \emph{tags} to uniquely identify threads \mbox{$\ptag{t}P$}. 
Uniqueness of tags is ensured by using the restriction operator and by 
only considering \emph{reachable} terms (Def.~\ref{def:reachable}).
Moreover, \respi\ extends \pic\ with three kinds of \emph{memories} $m$. 
An \emph{action memory} $\amem{\actEv{t_1}{\storedAct}{t_2}}{t_1'}{t_2'}$ 
stores an action event $\storedAct$ together with the tag $t_1$
of the active party of the action, the tag $t_2$ of the  
passive party, and the tags $t_1'$ and $t_2'$ of the new threads activated by the 
corresponding reduction.
An \emph{action event} records information 
necessary to revert each kind of interactions, which can be either a session initiation
$\initAct{a}{x}{y}{P}{Q}{s}$, 
a communication along an established session $\comAct{k}{e}{x}{P}{Q}$, 
or a branch selection $\selAct{k}{l_i}{P}{\branch{l_1}{P_1} \branchSep \ldots \branchSep \branch{l_n}{P_n}}$.
A \emph{choice memory} $\cmem{t}{\ifEv{e}{P}{Q}}{t'}$ stores a choice event 
together with the tag $t$ of the conditional choice and  $t'$ of the new activated thread.
The event $\ifEv{e}{P}{Q}$ records the evaluated expression $e$, 
and processes $P$ and $Q$ of the then- and else-branch, respectively.
A \emph{fork memory} $\fmem{\fev{t}{t_1}{t_2}}$ stores the tag $t$ of a splitting 
thread, of the form $\ptag{t}(P\mid Q)$, and the tags $t_1$ and $t_2$
of the new activated threads \mbox{$\ptag{t_1}P$} and \mbox{$\ptag{t_2}Q$};
these memories are analogous to \emph{connectors} in \cite{GLMT}. 
Threads and memories are composed by parallel composition and restriction operators.

Not all processes allowed by the syntax are semantically 
meaningful. In a general term, the history stored in the memories may not be
consistent, due to the use of non-unique tags or broken connections between continuation tags
within memories  and corresponding threads.  
For example, given the choice memory $\cmem{t}{\ifEv{e}{P}{Q}}{t'}$,
we have a broken connection when no thread tagged by $t'$ exists in the \respi\ process and 
no memory of the form $\amem{\actEv{t'}{\storedAct}{t_2}}{t_1'}{t_2'}$,
$\amem{\actEv{t_1}{\storedAct}{t'}}{t_1'}{t_2'}$,
$\cmem{t'}{\ifEv{e}{P_1}{P_2}}{t_1}$,
and $\fmem{\fev{t'}{t_1}{t_2}}$
exists.
Thus, as in~\cite{CristescuKV13}, to ensure history consistency we only consider \emph{reachable} 
processes, i.e.~processes obtained by means of forward and backward reductions from processes
with unique tags and no memory.

\begin{definition}[Reachable processes]\label{def:reachable}
The set of \emph{reachable} \respi\ processes is the closure 
under $\fwbwred$ (see below) of the set of terms, whose threads 
have distinct tags, generated by 
$
M\, ::= \, \ptag{t}P 
\ \ \mid\ \ 
\res{c}\, M
\ \ \mid\ \ 
M \mid N
\ \ \mid\ \ 
\nil
$.
\end{definition}

\paragraph{\respi\ semantics.}
The \respi\ operational semantics  is given in terms of a \emph{reduction relation} $\fwbwred$, given as the union 
of the forward and backward reduction relations. 
We report here, by way of examples, the forward and backward rules for session initiation (we require 
$s,\ce{s}$ fresh in $P_1$ and $P_2$ in the forward rule):
$$
	\begin{tabular}{@{}l@{}}
	$\ptag{t_1}\requestAct{a}{x}{P_1}\ \mid \ \ptag{t_2}\acceptAct{a}{y}{P_2}
	\ \fwred \  \res{s,t_1',t_2'}(
	\ptag{t_1'}P_1\subst{\ce{s}}{x} \mid \ptag{t_2'}P_2\subst{s}{y}
	\mid \amem{\actEv{t_1}{\initAct{a}{x}{y}{P_1}{P_2}{s}}{t_2}}{t_1'}{t_2'})$
	\\[.5cm]
	$\res{s,t_1',t_2'}(\ptag{t_1'}P \mid \ptag{t_2'}Q 
	\mid \amem{\actEv{t_1}{\initAct{a}{x}{y}{P_1}{P_2}{s}}{t_2}}{t_1'}{t_2'})
	\ \ \bwred\ \   \ptag{t_1}\requestAct{a}{x}{P_1}\ \mid \ \ptag{t_2}\acceptAct{a}{y}{P_2}$
	\end{tabular}
$$
When two parallel threads synchronise to establish a new session,
two fresh tags are created to uniquely identify the continuations. 
Moreover, all relevant information is stored in 
the action memory: 
the tag $t_1$ of the initiator (i.e., the thread executing a prefix of the form
$\bar{a}(\cdot)$), the tag $t_2$ of the thread executing the dual action, 
the tags $t_1'$ and $t_2'$ of their continuations, 
the shared channel $a$ used for the synchronisation, the replaced variables $x$ and $y$,
the generated session channel $s$, 
and the processes $P_1$ and $P_2$ to which substitutions are applied. 
All such information is exploited in the backward rule to revert this reduction.
In particular, the corresponding backward reduction is triggered by the
coexistence of the memory described above with two threads tagged 
$t_1'$ and $t_2'$, all of them within the scope of the session channel $s$
and tags $t_1'$ and $t_2'$ generated by the forward reduction (which, in fact, are 
removed by the backward one).
When considering reachable processes, due to tag uniqueness, processes $P$ and $Q$
coincide with $P_1\subst{\ce{s}}{x}$ and $P_2\subst{s}{y}$;
indeed, as registered in the memory, these latter processes have 
been tagged with $t_1'$ and $t_2'$ by the forward reduction. 
Therefore, the fact that two threads tagged with $t_1'$ and $t_2'$ are 
in parallel with the memory ensures that all actions possibly executed by 
the two continuations activated by the forward computation have been undone and, 
hence, we can safely undone the forward computation itself. 

\paragraph{Multiple providers scenario.}
\label{sec:example}
The scenario involving a client and two providers informally introduced in Section~\ref{sec:intro} 
is rendered in \respi\ as 
$
(\ptag{t_1}P_{client}  \ \mid \ \ptag{t_2}P_{provider1}  \ \mid \ \ptag{t_3}P_{provider2} )
$,
where the client process $P_{client}$ is 
\begin{center}
$
\begin{array}{l}
\requestPrefix{a_{login}}{x}.\,\send{x}{\mathsf{srv\_req}}.\,\receive{x}{y_{quote}}.\,
\ifPi\ accept(y_{quote})\ \thenPi\ \selectAct{x}{l_{acc}}{\,P_{acc}}\\
\hspace*{4.5cm}
\elsePi\ (\ifPi\ negotiate(y_{quote})\ \thenPi\ \selectAct{x}{l_{neg}}{\,P_{neg}}\ 
\elsePi\ \selectAct{x}{l_{rej}}{\,\inact})
\end{array}
$
\end{center}
while $P_{provider\,i}$ is 
$$
\acceptPrefix{a_{login}}{y}.\,\receive{y}{z_{req}}.\,\send{y}{quote_i(z_{req})}.\,
\branchAct{y}{\branch{l_{acc}}{Q_{acc}} \ \branchSep\ \branch{l_{neg}}{Q_{neg}} \ \branchSep\ \branch{l_{rej}}{\inact}}
$$

If the client contacts the first provider and accepts the proposed quote, 
the system evolves to
$$
M \, = 
\res{s,\ldots,t_1',t_2'}
(\,\ptag{t_1'}P_{acc}\msubst{\ssubst{\ce{s}}{x},\ssubst{\mathsf{quote}}{y_{quote}}}
\ \mid \
\ptag{t_2'}Q_{acc}\msubst{\ssubst{s}{y},\ssubst{\mathsf{srv\_req}}{z_{req}}} 
\ \mid \
m_1
\ \mid \
\ldots
\ \mid \
m_5
)
\mid \ P_{provider2}  
$$ 
where memories $m_i$ keep track of the computation history.
Now, if a problem occurs during the subsequent interactions, the computation can be reverted to allow 
the client to start a new session with (possibly) another provider:
$$
M \ \bwred^* \ptag{t_1}P_{client}  \ \mid \ \ptag{t_2}P_{provider1}  \ \mid \ \ptag{t_3}P_{provider2} 
$$

\paragraph{Properties of \respi.}
We show here that \respi\ enjoys standard properties of reversible calculi.

First, we demonstrate that \respi\ is a conservative extension of the (session-based) \pic.
In fact, as most reversible calculi, \respi\ is only a decoration of its host calculus. 
This decoration can be erased by means of the \emph{forgetful map} 
$\forgetMap$, mapping \respi\ terms into 
\pic\ ones by removing memories, tag annotations and tag restrictions. 
The following lemmas show that each forward reduction of a \respi\ process corresponds to 
a reduction of the corresponding \pic\ process and vice versa. 

\begin{lemma}\label{lem:correspondence}
Let $M$ and $N$ be two \respi\ processes. If $M \fwred N$ then $\forget{M} \red \forget{N}$.
\end{lemma}

\begin{lemma}\label{lem:correspondence_inv}
Let $P$ and $Q$ be two \pic\ processes. If $P \red Q$ then 
for any \respi\ process $M$ such that $\forget{M}=P$ there exists a \respi\ process
$N$ such that $\forget{N}=Q$ and $M \fwred N$.
\end{lemma}

Then, we show that \respi\ backward reductions are the inverse 
of the forward ones and vice versa.
\begin{lemma}[Loop lemma]\label{lemma:loop}
Let $M$ and $N$ be two reachable \respi\ processes. 
$M \fwred N$ if and only if $N \bwred M$.
\end{lemma}

We conclude with the \emph{causal consistency} result stating that two 
sequences of reductions (called \emph{traces} and ranged over by 
$\trace$), with the same initial state (\emph{coinitial}) and 
equivalent w.r.t. the standard notion of \emph{causal equivalence} 
($\causallyEq$), lead to the same final state (\emph{cofinal}). 
Thus, in this case, we can rollback to the initial state by 
reversing any of the two traces. 

\begin{theorem}
Let $\trace_1$ and $\trace_2$ be coinitial traces. Then, $\trace_1 \causallyEq \trace_2$
if and only if $\trace_1$ and $\trace_2$ are cofinal. 
\end{theorem}
\section{Discussion on a type discipline}\label{sec:types}
A question that should be answered before defining a static type discipline for a reversible 
calculus is ``\emph{Should we type check the processes stored in the memories?}''.
The question arises from the fact that we should be able to determine if any 
\respi\ process is well-typed or not.
In our case the answer is ``\emph{Yes}'', otherwise 
typability would not be preserved under reduction (i.e.,  Subject Reduction would not 
be satisfied).
It is indeed easy to define a \respi\ process (see \cite{TR}) containing a memory that, even if consistent,
triggers a backward reduction leading to an untypable term (by the type system 
defined in \cite{YoshidaV07} for the host calculus). 

One could wonder now if it is possible to type 
\respi\ processes in a na\"ive way by separately type checking the term
resulting from the application of $\forgetMap$ and each single memory, 
by using the type system in \cite{YoshidaV07}. 
For each memory we would check the term that has triggered the forward 
reduction generating the memory. 
In general, this approach does not work, because the term 
stored in a memory cannot be type checked in isolation
without taking into account its context. For example, 
consider a memory corresponding to a communication along a session
$s$ typable under typing
$\typing \,=\, \ce{s}:\outType{\intType}.\inactType \comp s:\inpType{\intType}.\inactType$ and
in parallel with $(\ptag{t_1}\send{\ce{s}}{1} \mid \ptag{t_2}\receive{s}{x})$.
The term resulting from the corresponding backward reduction is 
not typable, because the typings of its sub-terms  
are not composable (indeed, $\typing \comp \typing$ is not defined). 

Memory context can be considered by extending the type system 
in \cite{YoshidaV07} with rules that permits typing (processes stored in) memories 
and ignoring tag annotations and tag restrictions (see \cite{TR} for the definition of this type system).  
In this way, during type checking, typings of memories and threads must be composed by means of the rule for parallel composition. 
Thus, e.g., the \respi\ process mentioned above is, rightly, untypable. 
This type system properly works only on a simplified setting, 
which permits avoiding to deal with dependencies among memories and the threads 
outside memories, that could cause unwanted conflicts during type checking. 
Specifically, we consider the class of \respi\ processes that,
extending Def.~\ref{def:reachable}, are obtained by means of 
forward and backward reductions from 
processes with unique tags, no memory, no session initialised, no conditional choices and 
recursions at top-level, and no delegation. 
The characteristic of these processes is that, for each memory  
inside a process, there exists within the process an ancestor memory 
corresponding to the initialisation of the considered session. 
The type system checks only this latter kind of memories,
which significantly simplifies the theory. 

Coming back now to the multiple providers scenario,
we can verify that the initial process is well-typed. 
In particular, the channel $a_{login}$ can be typed by the shared channel type
$$
\sharedChanType{\,\inpType{\mathsf{Request}}.\,\outType{\mathsf{Quote}}.\,
\branchType{\branch{l_{acc}\!\!}{\!\!\alpha_{acc}}\,,\,\branch{l_{neg}\!\!}{\!\!\alpha_{neg}}\,,\,\branch{l_{rej}\!\!}{\!\!\inactType}}\,}
$$
where sorts $\mathsf{Request}$ and $\mathsf{Quote}$ are used to type requests and quotes,
respectively.
Let us consider now a scenario where the client wills to concurrently submit two different 
requests to the same provider, which would concurrently serve them. 
Consider in particular the following specification of the client
$$
\requestPrefix{a_{login}}{x}.\,(\,\send{x}{\mathsf{srv\_req\_1}}.\,P_{1} \ \mid \ \send{x}{\mathsf{srv\_req\_2}}.\,P_{2} \,)
$$
The new specification is clearly not well-typed, due to the use of 
parallel threads within the same session. This permits avoiding mixing up messages related to different 
requests and wrongly delivering them. In order to properly concurrently submit separate requests,
the client must instantiate separate sessions with the provider, one for each request.
\section{Concluding remarks}
\label{sec:conclusions}

To bring the benefits of reversible computing to structured communication-based programming,
we have defined a theoretical framework based on \pic\ that can be used as formal basis for studying
the interplay between (causal-consistent) reversibility and session-based structured interaction.  

The type discipline for \respi\ is still subject of study.  
In fact, the type system mentioned in Section~\ref{sec:types} is 
not completely satisfactory, because its use is limited to a restricted class of processes. 
To consider a broader class, 
an appropriate static type checking approach 
for memories has to be devised. For each memory, we would check 
a term composed of the threads stored in the memory and of a context composed of 
threads that have not been generated by the execution of the memory threads. 

Concerning the reversible calculus, we plan to investigate the definition of 
a syntactic characterisation of consistent terms, which statically enforces history consistency 
in memories (as in \cite{LaneseMS10}). 
It is worth noticing that the calculus is \emph{fully} reversible, i.e. 
backward computations are always enabled.
Full reversibility provides theoretical foundations for studying reversibility in session-based \pic, 
but it is not suitable for a practical use. In line with \cite{LaneseMSS11}, we plan to enrich the language with 
mechanisms to control reversibility.
Moreover, we intend to enrich the framework with an \emph{irreversible} action for committing 
the closure of sessions. 
In this way, computation would go backward and forward, allowing the parties to try different 
interactions, until the session is successfully completed.
For instance, the process $P_{acc}$ in our example could terminate 
by performing the irreversible action $\commit{x}$, which has to synchronise 
with action $\commit{y}$ in $Q_{acc}$. 
Differently from sessions terminated by $\inact$, a session terminated by a $\mathtt{commit}$ 
synchronisation is unbacktrackable. The irreversibility is due to the fact that no backward rule is defined to revert  this interaction. The type theory should be tailored to properly 
deal with this kind of session closure. 

As longer-term goals, we intend to apply the proposed approach to other session-based 
formalisms, which consider, e.g., asynchronous sessions and multiparty sessions. 
Moreover, we plan to investigate implementation issues that may arise when incorporating the 
approach into standard programming languages, in particular in case of a distributed 
setting.

\label{sect:bib}
\bibliographystyle{eptcs}
\bibliography{biblio}

\begin{thebibliography}{10}
\providecommand{\bibitemdeclare}[2]{}
\providecommand{\surnamestart}{}
\providecommand{\surnameend}{}
\providecommand{\urlprefix}{Available at }
\providecommand{\url}[1]{\texttt{#1}}
\providecommand{\href}[2]{\texttt{#2}}
\providecommand{\urlalt}[2]{\href{#1}{#2}}
\providecommand{\doi}[1]{doi:\urlalt{http://dx.doi.org/#1}{#1}}
\providecommand{\bibinfo}[2]{#2}

\bibitemdeclare{inproceedings}{CardelliL11}
\bibitem{CardelliL11}
\bibinfo{author}{Luca \surnamestart Cardelli\surnameend} \&
  \bibinfo{author}{Cosimo \surnamestart Laneve\surnameend}
  (\bibinfo{year}{2011}): \emph{\bibinfo{title}{Reversible structures}}.
\newblock In: {\sl \bibinfo{booktitle}{CMSB}}, \bibinfo{publisher}{ACM}, pp.
  \bibinfo{pages}{131--140}, \doi{10.1145/2037509.2037529}.

\bibitemdeclare{inproceedings}{CristescuKV13}
\bibitem{CristescuKV13}
\bibinfo{author}{I.~\surnamestart Cristescu\surnameend},
  \bibinfo{author}{J.~\surnamestart Krivine\surnameend} \&
  \bibinfo{author}{D.~\surnamestart Varacca\surnameend} (\bibinfo{year}{2013}):
  \emph{\bibinfo{title}{{A Compositional Semantics for the Reversible
  p-Calculus}}}.
\newblock In: {\sl \bibinfo{booktitle}{LICS}}, \bibinfo{publisher}{IEEE}, pp.
  \bibinfo{pages}{388--397}, \doi{10.1109/LICS.2013.45}.

\bibitemdeclare{inproceedings}{DanosK04}
\bibitem{DanosK04}
\bibinfo{author}{Vincent \surnamestart Danos\surnameend} \&
  \bibinfo{author}{Jean \surnamestart Krivine\surnameend}
  (\bibinfo{year}{2004}): \emph{\bibinfo{title}{Reversible Communicating
  Systems}}.
\newblock In: {\sl \bibinfo{booktitle}{CONCUR}}, {\sl \bibinfo{series}{LNCS}}
  \bibinfo{volume}{3170}, \bibinfo{publisher}{Springer}, pp.
  \bibinfo{pages}{292--307}, \doi{10.1007/978-3-540-28644-8\_19}.

\bibitemdeclare{article}{DanosK07a}
\bibitem{DanosK07a}
\bibinfo{author}{Vincent \surnamestart Danos\surnameend} \&
  \bibinfo{author}{Jean \surnamestart Krivine\surnameend}
  (\bibinfo{year}{2007}): \emph{\bibinfo{title}{{Formal Molecular Biology Done
  in CCS-R}}}.
\newblock {\sl \bibinfo{journal}{Electr. Notes Theor. Comput. Sci.}}
  \bibinfo{volume}{180}(\bibinfo{number}{3}), pp. \bibinfo{pages}{31--49},
  \doi{10.1016/j.entcs.2004.01.040}.

\bibitemdeclare{techreport}{GLMT}
\bibitem{GLMT}
\bibinfo{author}{E.~\surnamestart Giachino\surnameend},
  \bibinfo{author}{I.~\surnamestart Lanese\surnameend}, \bibinfo{author}{C.A.
  \surnamestart Mezzina\surnameend} \& \bibinfo{author}{F.~\surnamestart
  Tiezzi\surnameend} (\bibinfo{year}{2013}):
  \emph{\bibinfo{title}{{Causal-Consistent Reversibility in a Tuple-Based
  Language}}}.
\newblock \bibinfo{type}{Technical Report}.
\newblock
  \bibinfo{note}{\url{http://www.cs.unibo.it/~lanese/work/klaimrev-TR.pdf}}.

\bibitemdeclare{inproceedings}{LaneseMSS11}
\bibitem{LaneseMSS11}
\bibinfo{author}{I.~\surnamestart Lanese\surnameend}, \bibinfo{author}{C.A.
  \surnamestart Mezzina\surnameend}, \bibinfo{author}{A.~\surnamestart
  Schmitt\surnameend} \& \bibinfo{author}{J.~\surnamestart Stefani\surnameend}
  (\bibinfo{year}{2011}): \emph{\bibinfo{title}{{Controlling Reversibility in
  Higher-Order Pi}}}.
\newblock In: {\sl \bibinfo{booktitle}{CONCUR}}, {\sl \bibinfo{series}{LNCS}}
  \bibinfo{volume}{6901}, \bibinfo{publisher}{Springer}, pp.
  \bibinfo{pages}{297--311}, \doi{10.1007/978-3-642-23217-6\_20}.

\bibitemdeclare{inproceedings}{LaneseMS10}
\bibitem{LaneseMS10}
\bibinfo{author}{I.~\surnamestart Lanese\surnameend}, \bibinfo{author}{C.A.
  \surnamestart Mezzina\surnameend} \& \bibinfo{author}{J.~\surnamestart
  Stefani\surnameend} (\bibinfo{year}{2010}): \emph{\bibinfo{title}{Reversing
  Higher-Order Pi}}.
\newblock In: {\sl \bibinfo{booktitle}{CONCUR}}, {\sl \bibinfo{series}{LNCS}}
  \bibinfo{volume}{6269}, \bibinfo{publisher}{Springer}, pp.
  \bibinfo{pages}{478--493}, \doi{10.1007/978-3-642-15375-4\_33}.

\bibitemdeclare{article}{PhillipsU07}
\bibitem{PhillipsU07}
\bibinfo{author}{Iain C.~C. \surnamestart Phillips\surnameend} \&
  \bibinfo{author}{Irek \surnamestart Ulidowski\surnameend}
  (\bibinfo{year}{2007}): \emph{\bibinfo{title}{Reversing algebraic process
  calculi}}.
\newblock {\sl \bibinfo{journal}{J. Log. Algebr. Program.}}
  \bibinfo{volume}{73}(\bibinfo{number}{1-2}), pp. \bibinfo{pages}{70--96},
  \doi{10.1016/j.jlap.2006.11.002}.

\bibitemdeclare{techreport}{TR}
\bibitem{TR}
\bibinfo{author}{Francesco \surnamestart Tiezzi\surnameend} \&
  \bibinfo{author}{Nobuko \surnamestart Yoshida\surnameend}
  (\bibinfo{year}{2014}): \emph{\bibinfo{title}{{Towards Reversible
  Sessions}}}.
\newblock \bibinfo{type}{Technical Report}.
\newblock
  \bibinfo{note}{\url{http://cse.lab.imtlucca.it/~tiezzi/papers/places2014_ful%
l.pdf}}.

\bibitemdeclare{article}{YoshidaV07}
\bibitem{YoshidaV07}
\bibinfo{author}{N.~\surnamestart Yoshida\surnameend} \& \bibinfo{author}{V.T.
  \surnamestart Vasconcelos\surnameend} (\bibinfo{year}{2007}):
  \emph{\bibinfo{title}{{Language Primitives and Type Discipline for Structured
  Communication-Based Programming Revisited: Two Systems for Higher-Order
  Session Communication}}}.
\newblock {\sl \bibinfo{journal}{Electr. Notes Theor. Comput. Sci.}}
  \bibinfo{volume}{171}(\bibinfo{number}{4}), pp. \bibinfo{pages}{73--93},
  \doi{10.1016/j.entcs.2007.02.056}.

\end{thebibliography}

\end{document}